\documentclass[9pt,journal,compsoc]{IEEEtran}
\usepackage[nocompress]{cite}

\usepackage{color}

\usepackage[pdftex]{graphicx}
\graphicspath{{Figures/}}

\usepackage{amsmath}
\usepackage{listings}
\usepackage[justification=centering]{caption}

\usepackage{capt-of}
\makeatletter
\AtBeginDocument{%
	\providecommand*\ext@lstlisting{lol}%
	\renewcommand{\fnum@lstlisting}{\lstlistingname\nobreakspace\thelstlisting}
}
\makeatother

\newcommand\MYhyperrefoptions{bookmarks=true,bookmarksnumbered=true,
  pdfpagemode={UseOutlines},plainpages=false,pdfpagelabels=true,
  colorlinks=true,linkcolor={black},citecolor={black},urlcolor={black},
  pdfauthor={Luc Jaulmes},
  pdftitle={Memory Vulnerability: A Case for Delaying Error Reporting}
}
\ifCLASSINFOpdf
	\usepackage[\MYhyperrefoptions,pdftex]{hyperref}
\else
	\usepackage[\MYhyperrefoptions,breaklinks=true,dvips]{hyperref}
	\usepackage{breakurl}
\fi

\begin{document}

\title{Memory Vulnerability: A Case for Delaying Error Reporting}
\author{Luc~Jaulmes, Miquel~Moret\'o, Mateo~Valero,~\IEEEmembership{Member,~IEEE,} and~Marc~Casas%

\IEEEcompsocitemizethanks{\IEEEcompsocthanksitem All authors are with the Barcelona Supercomputing Center (BSC)
and Universidad Politecnica de Catalunya (UPC)
\protect\\
E-mail: luc.jaulmes@bsc.es}}

\IEEEtitleabstractindextext{%
  \begin{abstract}
To face future reliability challenges, it is necessary to quantify the risk of error in any part of a computing system.
To this goal, the Architectural Vulnerability Factor (AVF) has long been used for chips.
However, this metric is used for offline characterisation, which is inappropriate for memory.
We survey the literature and formalise one of the metrics used, the \textit{Memory Vulnerability Factor}, and extend it to take
into account false errors.
These are reported errors which would have no impact on the program if they were ignored.
We measure the \textit{False Error Aware MVF (FEA)} and related metrics precisely in a cycle-accurate simulator, and
compare them with the effects of injecting faults in a program's data, in native parallel runs.
Our findings show that MVF and FEA are the only two metrics that are safe to use at runtime, as they both consistently give an
upper bound on the probability of incorrect program outcome.
FEA gives a tighter bound than MVF, and is the metric that correlates best with the incorrect outcome probability of all
considered metrics.
  \end{abstract}
}

\maketitle
\IEEEdisplaynontitleabstractindextext
\maketitle

\section{Introduction}
\label{sec:introduction}
\IEEEPARstart{R}{eliability} is a major roadblock to successfully design ever smaller and more power efficient high performance
computers~\cite{Mitra14}.
Fault rates in transistors and DRAM cells keep increasing with their miniaturization, and a shift in paradigm to gain in energy
efficiency, such as using near-threshold voltage~\cite{Kaul12}, could decuplate the effects of soft errors.

In order to assess the reliability implications of system design decisions, Mukherjee et al.\ devised a metric: the Architectural
Vulnerability Factor (AVF)~\cite{Mukherjee03}, which allows to accurately model the error rate of a chip.
This metric quantifies whether a given bit matters for reliability, using the probability that a fault at this bit may cause an
error in the final outcome of a program.
Averaging this value per component for various workloads indicates which components require error mitigating techniques.
For example, a speculative component such as a branch predictor will never affect reliability, while an instruction decode unit
has a major impact on any program.

In memory subsystems however, all storage bits are uniformly protected with the same level of Error Correcting Code (ECC).
This means that bits in memory that will have no impact on the program are protected at the same expense than bits that will have
a major impact on the program's outcome.
This is because it is impossible to statically assess which storage bits are more likely to affect program reliability.
Indeed, as the placement of data in memory can change at every program execution, all bits in DRAM are considered interchangeable.
Thus, to generalize the notion of AVF to memory, it is necessary to use a metric dynamically, depending on the data stored at
a given bit, and on the way the program uses this data, at the scale of a single program execution.

In this paper, we formalize the \textit{Memory Vulnerability Factor (MVF)}, which is targeted at approximating the probability of
error in a program and has been used under different names in the literature.
We then extend this metric to take into account false errors, which are reported but would otherwise have no impact on the
program being run.
The MVF metric is dynamic, adapting to program behaviour, yet is not application-specific as it can be used for any program.
We measure it using a cycle-accurate simulator, as well as all previous generalisations of AVF to memory~\cite{Yu14,Luo14,Gupta18}.
We compare these metrics to the probability of an architecturally incorrect program outcome due to a fault in its data, obtained
from real-world fault injections.
These experiments demonstrate that the false error aware memory vulnerability metric correlates best with the probability of
an architecturally incorrect program outcome, and gives a consistent upper bound on this probability.

The paper is organised as follows: in Section~\ref{sec:metric_definition} we define the vulnerability metrics for memory, while
in Section~\ref{sec:related_work} we introduce the metrics used in the state of the art.
In Section~\ref{sec:methodology} we present our experimental setup and in Section~\ref{sec:eval} our results, before presenting
our concluding remarks in Section~\ref{sec:conclusion}.

\section{Metric Definition}
\label{sec:metric_definition}
The goal of quantifying memory vulnerability is to assess which bits matter to the outcome of a program.
In order to do this, we first present a classification of the possible outcomes of a program, then introduce the MVF metric,
whose scope is limited to memory, while taking into account program behaviour.
We then present the \textit{False Error Aware (FEA) MVF}, improving the MVF by analysing how false errors can in fact be ignored.

\subsection{Linking Program Outcome and Vulnerability}
\label{ssec:outcomes}

We categorise the outcome of a program as an Architecturally Correct Execution (ACE) when its execution is indistinguishable
from an execution without errors.
This includes executions with errors that are corrected, ignored, or benign (thus have no measurable impact on the final state).
Architecturally incorrect executions (un-ACE) may be due to a program that finishes running improperly (e.g. crash), stops making
progress, finishes but returns an incorrect result, or finishes and returns a correct result but having performed more work than
a non-faulty execution.
The goal of assessing vulnerability in memory is to know, dynamically, which bits have a higher likelihood of causing an un-ACE
outcome.

When encoding data with ECC, data bits are stored together with redundant bits, as an ECC codeword.
This makes computing a per-bit metric such as vulnerability slightly more complex.
For example a Single-Error-Correct Double-Error-Detect (SECDED) code means any single bit error in the codeword is
correctable~\cite{Hamming50}.
However, the bits in the codeword still affect the state of the program.
To reflect this, we attribute to an ECC codeword stored in memory the average vulnerability of all the bits in the unencoded data.
This allows to quantify the importance of the data that is encoded beyond the simple ECC strength considerations.
The properties of the AVF are also maintained: the structure's value is still the average of that of all its bits.
Thus, in order to assess the real impact of faults in data stored in memory, we need to measure the program outcomes when
injecting faults in the unencoded data bits.

\subsection{Memory is Vulnerable Before Being Loaded}
\label{ssec:vulndef}

As storage bits are interchangeable from one program execution to another, the vulnerability of a memory location is linked to
the data that is stored at that location, rather than to the physical DRAM cells themselves.
Therefore, we choose to use a definition of memory vulnerability that depends dynamically on the use of a given memory location
by a program.

To do this, we classify data residing in memory as either vulnerable or safe, based on the next memory access to this data.
Data may be read or overwritten.
We classify data as vulnerable when the next access to this data is a load, as any faults in this data will be propagated in the
memory request.
Conversely, data is safe when the next access to it is a store.
We then define the \emph{Memory Vulnerability Factor (MVF)} for each memory location as \textit{the fraction of time a memory
location contains data that will be read}.
That is, ${MVF = vulnerable~time / (safe~time + vulnerable~time)}$.

The core idea of this definition of the MVF is that only data that is loaded from memory can impact the program state, while
ignoring any effects outside memory.
Such effects can mask errors in data that is loaded in memory, for example if the data is not used by the program, has a
negligible impact, or has been loaded speculatively.

\subsection{Accounting for False Detected Uncorrected Errors}
\label{ssec:feadef}

The MVF is overestimating the probability of a bit affecting the program outcome, by considering data that is fetched only to be
overwritten as vulnerable.
Let us look at a write or a set of writes that spans a full ECC word (thus 8B for SECDED or 16B for ChipKill-level
ECC~\cite{Dell97}).
In a cache with a write-allocate policy, if these writes cause a cache miss, data will be fetched from memory.
Any miscorrected errors in this ECC word are masked by the new data being written.
Similarly, any uncorrectable errors would be masked as well, if they did not trigger an exception: these are called false errors.
Contrarily to benign errors, which affect the program in a negligible way, false errors are caused by faults in data that is not
consumed by the program.
Thus, in a system without ECC, or with an ECC scheme that has no Detected Uncorrected Errors (DUEs)~\cite{Abdoo96}, the
MVF incorrectly categorises that data as vulnerable, as it is based only on fetching the data.

A system with DUE can be enhanced to avoid causing exceptions on false errors, by simply delaying the reporting of any error
until it is actually consumed.
This is a common pattern in resilience, used for example to handle DUEs in memory discovered during scrubbing.
An application likely to access that erroneous data is not terminated preventively, but only whenever it attempts to access the
location of the error~\cite{Kleen10}.
Similarly at the architectural level, delaying a machine check exception due to an incorrect instruction allows to avoid
exceptions for instructions that do not affect correctness (no-ops, prefetches, etc.), or whose results are either not committed
or ultimately do not affect the program~\cite{Weaver04}.

Thus, we can write off false errors as not affecting the program state.
Accordingly, we differentiate between MVF, whose definition is given in Section~\ref{ssec:vulndef} and which limited to what
happens in memory, and \textit{False Error Aware MVF (FEA)}.
To compute the latter, we consider a memory location as safe not only when it is next accessed by a store, but also when it is
next accessed by a fill request whose contents will be overwritten.
This slight difference in what classifies as a vulnerable memory access does not affect the metrics' mathematical properties.

\subsection{Vulnerability under Transient Fault Models}

Both the MVF and FEA metrics relate naturally to the probability of consuming a fault, under the common hypotheses for transient
fault models.

If we suppose that faults happen randomly and independently of one another, we can model them using an exponential model, with
$\lambda$ the average fault arrival rate.
Formally, we define for any location in memory $S$ and $U$, the sets of safe and unsafe accesses to that location (i.e.\
respectively stores and loads for the MVF), and $t_a$ and $f_a$, the time and the number of faults consumed by any access $a$
respectively.
Since safe accesses overwrite faults, we have ${a \in S \Rightarrow f_a = 0}$.
For unsafe accesses, we consider the period $p_a$ before an access $a$.
This lasts ${p_a = t_a - t_{prev(a)}}$, where ${prev(a) = \max\{b \in S \cup U | t_b < t_a\}}$.
Thus for unsafe accesses, we have ${P(f_a > 0) = 1 - P(f_a = 0) = 1 - e^{-\lambda p_a}}$, hence the overall probability of
consuming a fault is ${P(F) = \sum_{u \in U} (1 - e^{-\lambda p_u})}$.
If we reasonably assume that faults are rare, i.e.\ that the program execution time $T$ is such that ${\lambda T \ll 1}$,
then we can approximate ${P(F) \approx \lambda \sum_{u \in U} p_u}$, which is the total time spent before unsafe accesses
multiplied by ${\lambda}$.

The vulnerability $V$ is the fraction of time a memory location contains data that is unsafe, thus with the same notations, ${V =
\frac{1}{T}\sum_{u \in U} p_u}$.
We then have ${P(F) \approx \lambda TV}$, which confirms the intuition of the vulnerability $V$ being proportional to the
probability of consuming a fault in memory.

\section{Related Work Gauging Memory Vulnerability}
\label{sec:related_work}
Previous work has studied the variable vulnerability of data in memory using metrics similarly inspired by the AVF.

Yu et al.\ define the \textit{Data Vulnerability Factor (DVF)}~\cite{Yu14} per data structure $d$, defined as the multiplication
of the structure's size $S_d$, the program execution time $T$, the number of accesses to this structure in memory $N_{ha}$, and
the overall fault rate $FIT$: ${DVF_d = FIT \cdot T \cdot S_d \cdot N_{ha}}$.
They then use mathematical \linebreak models to compute the DVF based on memory access patterns.

Luo et al.\ use the \textit{safe ratio}, which is the fraction of time that data resides in memory before being
overwritten~\cite{Luo14}.
This is the same as the MVF, except that it chooses to quantify the opposite of vulnerability.
The MVF and the safe ratio $sr$ are related by ${MVF + sr = 1}$.
Luo et al.\ do not inject faults in native runs, instead using a debugger.

Gupta et al.~\cite{Gupta18}\ use two metrics, initially measuring the MVF,
defined as ``the average duration that data is stored in memory before being loaded''.
They then use a proxy metric, which is the ratio of stores ($ST$) to loads ($LD$), thus $ST/LD$, for the purpose of their runtime
page-placement algorithm.
The analysis of the impact of faults is never done with the real program outcome, as the experiments are run inside FaultSim, a
Monte-Carlo model-based simulator, where loading an uncorrectable or undetected error is equated with program failure.

\section{Methodology}
\label{sec:methodology}
To compare the vulnerability metrics with the probability of un-ACE outcomes in the event of faults, we examine outcomes when
injecting faults in native parallel runs of the Conjugate Gradient (CG), and measure vulnerability ratings
precisely using a cycle-accurate simulation infrastructure.
We first explain the simulator setup, and then detail the fault injection experiments before presenting the benchmark, CG.

\subsection{Measuring the Memory Vulnerability}
\label{ssec:tasksim}

To be able to gather enough information about when data reaches or is fetched from memory, we use a cycle-accurate simulator.
We extend TaskSim~\cite{Rico11,Rico12}, a task-trace based multicore simulator, to compute the exact memory vulnerability ratings
of data.
Its infrastructure relies on task-based execution models to generate detailed traces for each task, including the basic blocks
that are executed and memory addresses that are accessed.
TaskSim's multicore architecture simulator then simulates parallel runs in detail by fetching and simulating all instructions,
using a simple core model and a full memory hierarchy.
The simulator also relies on a real runtime system, to schedule the tasks across the simulated hardware.

To compute the various vulnerability metrics, we capture all loads and stores and the time at which they reach main memory.
We then update at each access the necessary counters per memory location: time before stores, time before loads, FEA safe time
(before fill requests whose contents will be overwritten), and number of loads and stores.
From this data, we compute the fraction of time that each location is vulnerable, the DVF and the load-to-store ratio.
We only update these counters during the Region Of Interest (ROI), thus while the solver is running.
We compute all metrics at a 64 bit granularity, which is the granularity used for SECDED and a subset of the granularity commonly
used in ChipKill-level ECC.

We trace applications on an Intel x86\_64 Xeon E5-2670 and simulate a multicore architecture whose configuration mirrors the Xeon
E5-2670's characteristics.
It consists of 8 cores running at a frequency of 2.6GHz, each with a reorder buffer of 168 entries, and one thread per core.
The memory hierarchy's parameters are summed up in Table~\ref{tab:tsparams}.
All cache levels have 64B lines, write-back and write-allocate policies, are non inclusive,
and track outstanding misses in Miss Status Handling Registers (MSHRs).
In the simple model used for main memory, every request has the same latency and returns a full cache line.

\begin{table}[bt]
	\addtolength\tabcolsep{-1.5pt}\centering
	\caption{TaskSim cache and memory parameters}
	\label{tab:tsparams}
	\begin{tabular}{r|lllll}
		\hline
		cache & shared & assoc. &  size & latency & MSHRs \\
		\hline
		L1D & private & 8-way  & 32kB   & 4 cycles  & 32 \\
		L2  & private & 8-way  & 256kB  & 12 cycles & 32 \\
		L3  & shared  & 16-way & 20MB   & 28 cycles & 128 \\
		\hline
		memory & \multicolumn{2}{c}{max bandwidth} & size & \multicolumn{2}{l}{latency} \\
		\hline
		 & \multicolumn{2}{c}{16GB/s} & 32GB & \multicolumn{2}{l}{155 cycles}
	\end{tabular}
\end{table}

\subsection{Measuring the Program Outcome}
\label{ssec:injection}

To analyse and validate the memory vulnerability metrics, they need to be compared against the outcomes of injecting faults in
the memory of the program.
Thus, we inject faults in native parallel runs on a real system, using the same Xeon E5-2670 as used for the simulation
infrastructure.

Faults are injected using a separate thread, at a uniformly random point in the targeted application-level data, and at a
uniformly random time during the ROI.
The injector thread sleeps for the selected amount of time, then injects the single event by flipping the selected bit.
Each experiment consists of a single fault injection, and each experimental campaign consists of at least 6500 parallel CG runs
per data structure.

The prologue, which consists of generating the program's input, and the epilogue, which consists of verifying the program's
output, are not considered in this analysis.
The program runs until it finishes abnormally or until completion, in which case it verifies the validity of the solution it
found.
This verification is always done against the unmodified input data.
If the program performs more work (in this case, more iterations) than the ACE baseline, we classify it as un-ACE regardless of
the validity of the solution.
If the program does not finish within $10\times$ the baseline (ACE) execution time, we classify its execution as un-ACE and
interrupt it.

\subsection{The Conjugate Gradient Benchmark}
\label{ssec:cg}

\begin{figure}[bt]
	\centering%
	\captionof{lstlisting}{Conjugate Gradient (CG) pseudo code}
	\begin{minipage}[t]{0.85\linewidth}
		\begin{lstlisting}[label=code:cg,mathescape]
$\epsilon_{old} \Leftarrow +\infty$, $d' \Leftarrow 0$
for $t$ in $0..t_{max}$:
	$g \Leftarrow b - A x$ if $t \equiv 0\ (\text{mod}\ 50)$ else $g - \alpha q$
	$\epsilon \Leftarrow ||g||^2$
	if $\epsilon < tol$:break
	$\beta \Leftarrow \epsilon / \epsilon_{old}$
	$d \Leftarrow \beta d' + g$
	$q \Leftarrow Ad$
	$\alpha \Leftarrow \epsilon / \langle q,d \rangle$
	$x \Leftarrow x + \alpha d$
	$\epsilon_{old} \Leftarrow \epsilon$
	swap($d$,$d'$)
		\end{lstlisting}
	\end{minipage}%
\end{figure}

To perform our experiments, we use the Conjugate Gradient (CG) as a benchmark, whose pseudo-code is presented in
Listing~\ref{code:cg}~\cite{Shewchuk94}.
CG solves $Ax = b$ for $x$, where $A$ is a sparse symmetric positive definite matrix.
$x$, $g$, $d$, $d'$, and $q$ are vectors, and $\epsilon$, $\epsilon_{old}$, $\alpha$ and $\beta$ are scalars.
The matrix A is stored in memory in compressed sparse row format.
Thus, we refer to it as 3 separate data structures: the rows $Ar$, columns $Ac$, and values $Av$.
We solve the 3D Poisson's equation discretised with a 27 point stencil, and a size of $64^3$ rows.

We implement CG using OmpSs, a task-based dataflow programming model, with tasks each generating one block of each of the
vectors or sums (in the case of reductions)~\cite{Jaulmes15}.
We use two copies of $d$ and swap their pointers (line~15) to allow delaying tasks that depend on one copy, such as ${x
\Leftarrow x + \alpha d}$.
This allows the runtime to ignore the false dependency due to overwriting $d$, and to overlap $x$'s update with operations that
incur load imbalance such as $\langle d, q\rangle$, thus improving scalability.

\section{Evaluation}
\label{sec:eval}
In this section, we compare the probability of un-ACE outcomes when injecting faults against the MVF, FEA, and other
metrics used by related work.

\subsection{Memory Vulnerability Metric Validation}
\label{ssec:validation}

\begin{figure}[bt]
	\centering
	\includegraphics[width=\linewidth]{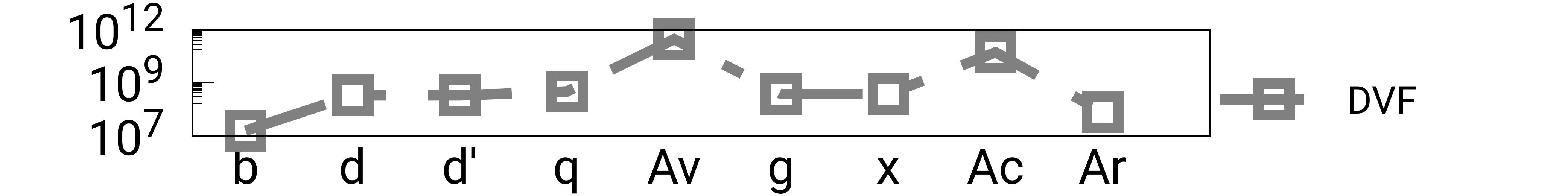}
	\includegraphics[width=\linewidth]{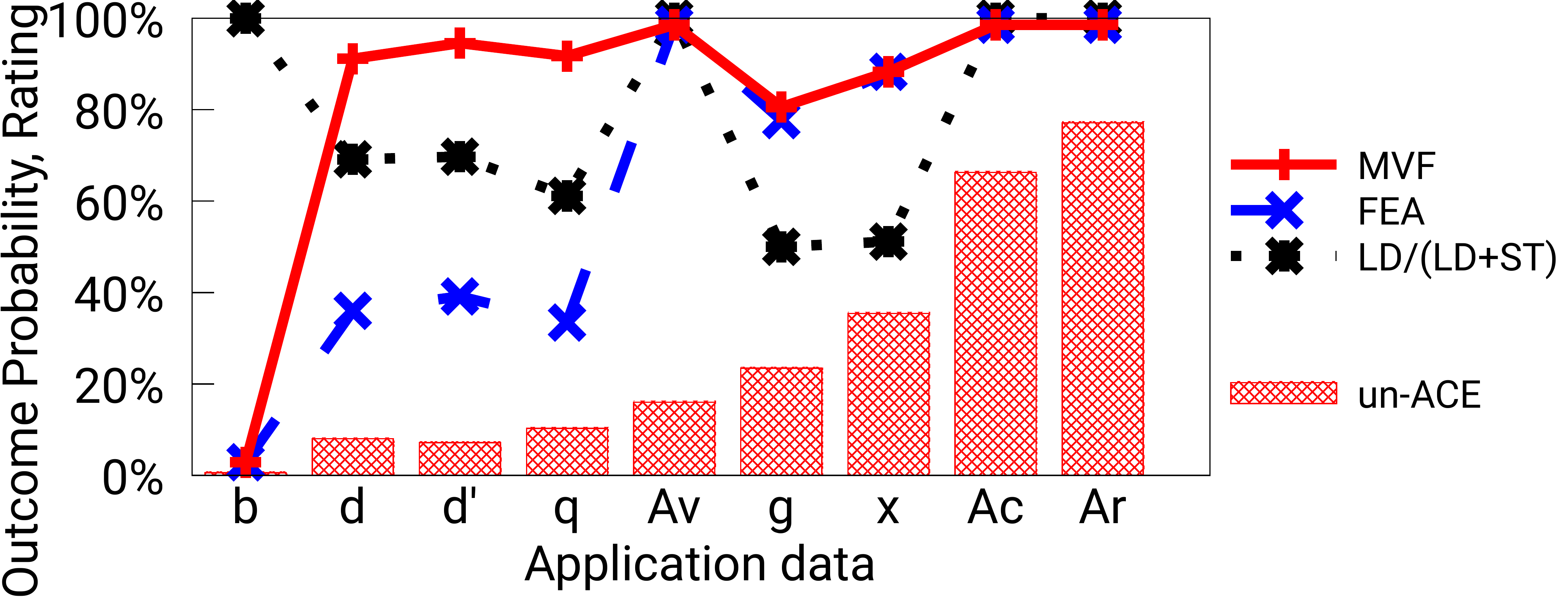}
	\caption{ACE frequency when injecting faults in real runs, and simulation-based vulnerability ratings per CG data structure}
	\label{fig:vect_bitflip}
\end{figure}

The results of fault injections in real runs, categorising the outcomes as architecturally correct (ACE) or not (un-ACE), are
presented as bars in Figure~\ref{fig:vect_bitflip}, while the various vulnerability ratings obtained from cycle-accurate
simulations are presented as lines.
The bars represent the probability of an un-ACE outcome, and are obtained from fault injections.
Data structures are sorted in increasing un-ACE probability, and confidence intervals are not displayed, as they are too small.
The largest interval is for $g$, whose probability of suffering an un-ACE outcome is in $[22.0\%, 24.6\%]$ with 99\% confidence.
The vulnerability metrics, displayed as lines, are the MVF, defined in Section~\ref{ssec:vulndef} and similar to the \emph{safe
ratio}~\cite{Luo14}, the FEA, defined in Section~\ref{ssec:feadef}, the load-to-store ratio~\cite{Gupta18}, and finally the
DVF~\cite{Yu14} in the separate graph above.

The probabilities of un-ACE outcomes are consistent with the memory access pattern of the CG
algorithm for each data structure, taking into account that the solver converges in 27 iterations for this matrix and that most
of the time of one iteration is spent computing the matrix-vector multiplication.
The $b$ vector is only used for the initial computation of the residual, and the data is never read afterwards, hence with a
vulnerability of $2.9\%$.
Since $Ac$ and $Ar$ are arrays of indexes, it is expected that faults in these data structures may cause crashes, in the form of
segmentation faults.
All other data consists of floating point values.
When injecting faults in these structures, un-ACE program outcomes are almost never crashes, but instead incorrect or slower
executions.
Furthermore, we can note that these data structures generally have lower un-ACE outcomes than the integer ones.
While this is due in part to the way data is accessed by the program, it is also inherent to floating point data, as shown by
the difference between $Ac$ and $Av$, which both have the exact same access pattern.
Errors in lower-order bits of floating point data likely have a negligible effect on the solver, as their relative significance is
small.

The various vulnerability metrics are presented as lines in Figure~\ref{fig:vect_bitflip}.
Comparing the MVF (unbroken line) and FEA (dashed line) to the un-ACE frequency, we can see that both metrics provide an upper
bound on this probability.
The FEA gives a much tighter bound for the $d$, $d'$ and $q$ vectors.
This is due to these vectors being overwritten and not updated, as can be seen in Listing~\ref{code:cg}.
Thus, false errors would be overwritten approximately 50\% of the time: after the vector is read and before it is computed again.
On the other hand, $g$ and $x$ are updated, thus with a read-modified-write sequence for each element.
Accordingly, both MVF and FEA evaluate these vectors in the same way.
Finally, $Av$, $Ac$, and $Ar$ are only ever read, and $b$ is only read once at the beginning of the execution, thus mostly
contains inconsequential data.
Overall, FEA correlates better with the un-ACE probability.

As the load-to-store ratio $LD/ST$~\cite{Gupta18} is unbounded ($b$'s rating is $+\infty$ for example) we normalize it as
${LD/(LD + ST)}$, and present this value as the dotted line in Figure~\ref{fig:vect_bitflip}.
This transformation maintains the relative order of the ratings' values while bringing them back in the interval $[0, 1]$.
We see several problems with this load-to-store ratio: $b$, which is mostly inconsequential, is rated equally vulnerable as $Ar$,
and both $x$ and $g$ have the lowest vulnerability ratings while being the third and fourth most vulnerable vectors, where faults
cause un-ACE results with probabilities of $25\%$ and $46\%$ respectively.
Furthermore, the load-to-store ratio over-estimates but also under-estimates the vulnerability of considerable chunks of data,
and can thus not be safely used as an upper bound on failure probability.
For example, the first parts of vector $q$ to be computed are very vulnerable, and the last ones very safe, as computing this
vector is an operation that takes a long time.
However, the load-to-store ratio rates the whole vector with the same vulnerability, by considering only the number of accesses.

The DVF metric~\cite{Yu14} is not bounded either and its values are high and very spread out, hence we display them on a log
scale at the top of Figure~\ref{fig:vect_bitflip}.
The main factor impacting the DVF of a data structure is its size, which causes $Ac$ and $Av$ to have the highest values.
The most critical data, $Ar$, is rated as the second safest data structure by the DVF.
The other values are very close together, in increasing order: $d$ and $d'$, followed by $g$ and $x$, and then $q$.
Overall this correlates poorly with the probability of un-ACE outcomes, and the fact that the metric is not bounded makes it
harder to use at runtime.
Indeed, the metric only has meaning when comparing values relative to each other, making it impossible to set thresholds on DVF
values for example.

\section{Conclusion}
\label{sec:conclusion}
A number of metrics aim at quantifying the risk associated with encountering an error in data in memory.
Comparing these metrics with the likelihood of architecturally incorrect executions due to a fault in memory indicates that
the \textit{False Error Aware Memory Vulnerability Factor (FEA)}, which is introduced in this paper, is the most accurate one.
This can be explained by the fact that it takes into account timing effects, as opposed to DVF or store-to-load ratios, and
the fact that overwriting data that has not been accessed in a while may cause false errors, which can be ignored.
This work also highlights that hardware support to track false errors in the cache hierarchy would likely significantly decrease
error rates due to DUE in memory.
Furthermore, this work opens the door to runtime-level optimizations that can now accurately model the risk associated with any
given data.

\ifCLASSOPTIONcompsoc
\section*{Acknowledgements} 
\else
\section*{Acknowledgement} 
\fi

This work has been supported by the RoMoL ERC Advanced Grant (GA 321253), by the European HiPEAC Network of Excellence,
by the Spanish Ministry of Economy and Competitiveness (contract TIN2015-65316-P), by the Generalitat de Catalunya
(contracts 2014-SGR-1051 and 2014-SGR-1272) and by the European Union’s Horizon 2020 research and innovation programme
(grant agreements 671697 and 779877).
L.\ Jaulmes has been partially supported by the Spanish Ministry of Education, Culture and Sports under grant FPU2013/06982.
M.\ Moret\'o has been partially supported by the Spanish Ministry of Economy, Industry and Competitiveness under Ram\'on y Cajal fellowship number RYC-2016-21104.

The authors would like to thank Francesc Mart\'inez Palau for his precious help and support with the TaskSim infrastructure.

\bibliographystyle{ieeetr}
\bibliography{IEEEabrv,references}

\end{document}